\documentclass[runningheads]{llncs}
\usepackage{listings}
\usepackage{xcolor}
\usepackage[hidelinks]{hyperref}
\lstdefinelanguage{json}{
    basicstyle=\ttfamily\scriptsize,
    numbers=left,
    numberstyle=\tiny\color{gray},
    stepnumber=1,
    numbersep=5pt,
    showstringspaces=false,
    breaklines=true,
    frame=single,
    backgroundcolor=\color{white},
    keywordstyle=\color{blue},
    stringstyle=\color{red},
    commentstyle=\color{gray},
    morecomment=[l][\color{magenta}]{//},
    literate=
     *{0}{{{\color{blue}0}}}{1}
      {1}{{{\color{blue}1}}}{1}
      {2}{{{\color{blue}2}}}{1}
      {3}{{{\color{blue}3}}}{1}
      {4}{{{\color{blue}4}}}{1}
      {5}{{{\color{blue}5}}}{1}
      {6}{{{\color{blue}6}}}{1}
      {7}{{{\color{blue}7}}}{1}
      {8}{{{\color{blue}8}}}{1}
      {9}{{{\color{blue}9}}}{1}
      {:}{{{\color{black}:}}}{1}
      {,}{{{\color{black},}}}{1}
      {"}{{{\color{red}"}}}{1},
}

\usepackage[T1]{fontenc}
\authorrunning{~}
\usepackage{graphicx}
\usepackage{hyperref}
\usepackage{soul, color}
\usepackage{enumitem}
\usepackage{float}
\begin{document}
\title{Cloud-Based Interoperability in Residential Energy Systems}
%
%
\author{Darren Leniston\inst{1} \and
David Ryan\inst{1} \and
Ammar Malik\inst{2} \and 
Jack Jackman\inst{1} \and
Terence O'Donnell\inst{2}}
%
%
\institute{Walton Institute, South East Technological University, Waterford, Ireland \and
School of Electrical \& Electronic Engineering, University College Dublin, Ireland}
\maketitle              
\begin{abstract}
As distributed energy resources (DERs) such as solar PV, batteries and electric vehicles become increasingly prevalent at the edge, maintaining grid stability requires advanced monitoring and control mechanisms. This paper presents a scalable smart grid gateway architecture that enables interoperability between Modbus-based inverters and IEEE 2030.5 cloud-based control systems. The proposed solution leverages Azure cloud services and edge-computing gateway devices to support dynamic configuration, telemetry ingestion, remote control and Volt-VAR Curve deployment. A microservice-based architecture ensures flexibility and scalability across diverse deployment scenarios, including both gateway-mediated and direct-to-cloud device communication. Results demonstrate the successful mapping of a Fronius Primo inverter's Modbus registers to IEEE 2030.5-compliant telemetry and control functions. Additionally, we evaluate real-time VVC updates and their impact on local voltage regulation, showcasing dynamic cloud-to-edge control with minimal latency. This work highlights the potential of virtualised, standards-based control infrastructures to support DER integration and active grid participation, while remaining adaptable to evolving smart grid architectures.

\end{abstract}
\section{Introduction}
\noindent As the number of renewable energy sources (RES) increases in tandem with the push from both National and European policymakers to mitigate the cumulative impacts of climate change, the energy sector is undergoing significant transformations to align with initiatives like the European Green Deal.
\noindent The research described in this paper is designed to support this transition by developing a platform to enable Smart Grid (SG) control architectures. This platform provides an example of how Distribution System Operators (DSO) and other stakeholders in the energy sector can effectively integrate SG devices for compliance with data and communication standards. To further promote the standardisation of SG devices, the research has developed an SG gateway device. 
\noindent This device facilitates communications with SG components such as smart inverters while managing telemetry data and control command execution. Built in accordance with the IEEE 2030.5 and Irish DSO network standards and protocols, it will enhance both the system and asset owners' visibility and control over their resources, ensuring interoperability between SG devices\cite{Johnson2021}.
\noindent Additionally, the research has implemented a cloud platform architecture to support bi-directional communication with gateway devices in the field. This cloud platform enables intelligent deployment, management, and monitoring of gateway devices in compliance with the requirements of the IEEE 2030.5 standard in terms of interfaces and cybersecurity. 

\section{Background}
\subsection{Remote Configuration in the Smart Grid} 
\noindent As the structure of the energy grid becomes increasingly distributed, utilities are implementing and leveraging the capabilities of edge computing hardware to enable monitoring and fine-grained control functionality. In this context, the ability to remotely configure these edge devices is crucial, ensuring the necessary level of flexibility, resilience, and efficiency required in modern distribution systems. Traditionally, applying configuration and firmware updates to smart meters, RTUs (Remote Terminal Units), and other assets located at the network edge required manual intervention, often involving physical access to the device itself \cite{Lekbich2018}. To address these challenges, modern SG systems rely on automated OTA (over-the-air) configuration techniques that leverage cloud computing capabilities, software-defined networking (SDN), and increasingly on AI and ML-based technologies.Furthermore, zero-touch-provisioning (ZTP), a paradigm leveraged in the research described in this paper, has further streamlined remote configuration by enabling edge devices to automatically retrieve and apply configuration settings upon deployment \cite{Demchenko2016}. This eliminates the need for manual setup and reduces operational costs, particularly in large-scale grid roll outs. Cloud-based device management platforms play a vital role in orchestrating these processes, allowing utilities to remotely monitor, configure, and troubleshoot devices in real time \cite{Demchenko2016}. AI and machine learning techniques can be utilised to further enhance these capabilities by predicting configuration failures, optimising parameter settings, and identifying anomalies that may indicate security threats \cite{Hazra2024}.

\subsection{Smart Grid Standardisation and Interoperability}
\noindent SG interoperability is a key factor for integrating Distributed Energy Resources (DERs) into modern energy grids. An assortment of standards facilitate seamless communication among diverse systems, enhancing grid management, energy distribution and reliability. These standards support the integration of Renewable Energy Resources (RES), enable demand response capabilities and support Advanced Metering Infrastructure (AMI). Communication standards are particularly critical, enabling the remote configuration and device management across heterogenous infrastructures. For example, IEC 61850 supports substation automation and real-time data exchange, while OpenADR enables dynamic demand response by adjusting devices based on grid conditions \cite{Ninagawa2015}. The IEEE 2030.5 standard, part of the SG Interoperability Series, governs utility interaction with consumer devices via internet protocols such as TCP/IP and a RESTful architecture leveraging XML. It is comprised of "Function Sets" covering DERs, EV support, billing and demand response. The standard utlises elements from IEC 61968 and supports multiple deployment models, including in-home and cloud-based setups \cite{ieee2030.5}. The Common Information Model (CIM) provides a standardised framework for representing power systems in both UML and XML \cite{crimmins2022}. The standard includes IEC 61970 (energy management), IEC 61968 (utility domain integration) and IEC 62325 (market communication) with the aim of promoting interoperability and efficiency \cite{crimmins2022}. SAREF (Smart Applicances REFerence) is a standard which provides an onotological view for aligning IoT assests across domains. Its modular structure allows for domain-specific extensions \cite{haghgoo2020}, such as SAREF4ENER which models DER flexibility for demand response, in addition to SAREF4GRID which standardises SG IoT data. 

\subsection{Digital Twins in the Smart Grid}
\noindent A digital twin serves as a virtual representation of a physical asset or system that updates in real-time using data collected from the edge \cite{Baek2022}. In smart grids, digital twins enhance monitoring, simulation and optimisation across power generation, transmission, distribution and consumption contexts, thereby improving efficiency, resilience and decision-making capabilities \cite{Lu2021}. A key benefit is predictive maintenance, where real-time data from assets such as transformers enables early fault detection, reducing downtime and associated costs. Digital twins also support contingency analysis, aiding operators in mitigating the impacts from extreme weather or load fluctuations \cite{Balijepalli2021}. Their implementation depends on IoT, machine learning and cloud/edge computing, where IoT sensors supply granular asset data, cloud platforms handle large-scale simulations and edge computing ensures low-latency responses. Interoperability with physical systems is supported by standards such as CIM, IEEE 2030.5 and IEC 61850 \cite{Balijepalli2021}. Extending beyond operational optimisation, digital twins also aid in real-time control, energy market simulations, DER integration and demand response testing. They also bolster cybersecurity by simulating attack scenarios. However, challenges such as high costs, data privacy and governance hinder broader adoption and research is ongoing to mitigate theses concerns by explore solutions such as blockchain-based secure data sharing \cite{Koeva2024}.

\section{Use Cases}
\subsection{Mapping of Inverter Communications to Smart Grid Standards}
\begin{figure}
    \centering
    \includegraphics[width=1\linewidth]{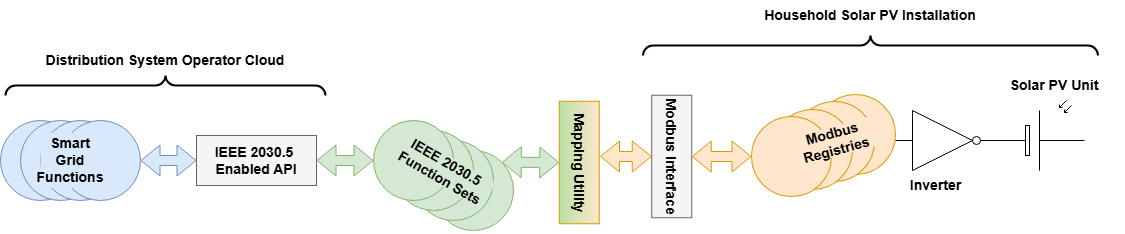}
    \caption{Solar PV to Smart Grid Communications mapping}
    \label{fig:pv2cloud_mapping}
\end{figure}

As protocols such as IEEE 2030.5 and SunSpec Modbus gain traction among DSOs, enhancing edge interoperability has become increasingly vital. The proliferation of residential solar PV systems for example, coupled with the heterogeneity of inverter technolocies, both compliant and non-compliant with SunSpec, challenge the DSOs' ability to maintain grid observability and stability. Consequently, new use cases are emerging to standardise telemetry and control interfaces across diverse devices. The use case described here addresses the mapping of Modbus registers to IEEE 2030.5 function sets for edge-to-cloud communication. Modbus, a widely used industrial protocol, follows a request/response model, typicall over RS485 or TCP/IP. It leverages four key data types, discrete inputs, coils, input registers and holding registers, to enable access to sensor data and control commands. A suitable smart inverter must support Modbus over TCP/IP and RS485 in addition to offering a set of registers with defined address spaces, data types and control functions. However, register configurations vary significantly be vendor, complicating interoperability and potentially causing communication failures with upstream platforms when data formats or mappings are inconsistent. This use case demonstrates how aligning Modbus register functions with IEEE 2030.5 function sets can establish a unified data layer between inverter telemetry/control systems and cloud-based grid management platforms, enabling more reliable integration and household participation in DSO-led grid balancing efforts. Figure \ref{fig:pv2cloud_mapping} illustrates this mapping, where SG functions are exposed through an IEEE 2030.5 API, linking a PV deployment to an IEEE 2030.5 compliant system while utilising Modbus registers to interface with inverter functions.

\subsection{Semi Autonomous DER Control}
\begin{figure}
    \centering
    \includegraphics[width=1\linewidth]{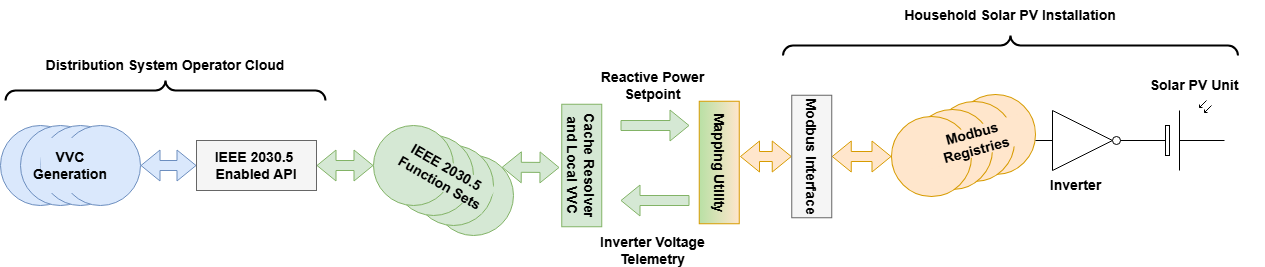}
    \caption{Cloud to Inverter VVC Transfer}
    \label{fig:vvc_to_inverter}
\end{figure}

As smart grids evolve, consumers increasingly act as prosumers, contributing grid services via DERs such as solar PV, batteries and Vehicle-to-Grid (V2G) enabled EVs. To prevent instability from widespread DER adoption, for example voltage fluctuations, frequency imbalances and reverse power flows, smart inverter-level control is essential. Modern inverters support Volt/VAR and Volt/Watt control, enabling voltage regulation in real-time. These capabilities, combined with bi-directional communication via ADMS and DERMS, allow dynamic adjustments that enhance grid resilience. Unlike simple curtailment, Volt-VAR Curves (VVCs) dynamically adjust reactive power in response to grid conditions. VVCs are particularly valuable in high DER or EV penetration areas, mitigating voltage fluctuations cause by intermittent generation and bidirectional energy flow, ensuring stability without reducing active power exports. The use case described explores the cloud-based generation of VVCs for clusters of DERs, and their delegation to inverter-level control. VVCs are scheduled with grid demand and capacity in mind, requiring systems that can distribute updates to edge devices. While some inverters can directly implement curves, others rely on receiving reactive power setpoints, introducing complexity in scaling across diverse devices. Figure \ref{fig:vvc_to_inverter} illustrates generation of the VVC in the DSO's cloud environment and transmission to the household DER installation for actuation, whether that be the inference of a reactive power set-point, as is in this case, or the autonomous programming of the VVC on the inverter. 

\section{Proposed Remote Configuration System for Interoperability and Standards Management}

\subsection{Edge Device Design}
The concept for an SG gateway is based on the IoT gateway model, acting as a central hub for communication with devices such as PV inverters and smart meters. These gateways and their functionality are increasingly critical across both distribution and transmission level, facilitating large-scale data retrieval and integration of diverse platforms across LV, MV and HV layers. Depending on their capabilities, SG gateways also support edge computing, enabling local data processing and aggregation to reduce network traffic and latency. Additionally, their customisable software and firmware with the capaibility of remote updating supports multiple use cases. Unlike more general IoT devices, SG devices require strict compliance with communication standards such as IEEE 2030.5 and (SunSpec) Modbus, which define common parameters for DER monitoring and control [1]. Thus, any SG gateway must support SunSpec modbus at minimum, in addition with other necessary protocols. Hardware requirements for such a gateway include RS485 ports for serial connections, wireless/ethernet for internet-based comms and/or cellular modules for remote connectivity. Gateways must by physically secure to protect data and grid assets and robust enough for deployment in the field. Typically running Unix-based systems, they support containerised microservices for flexible deployment and often integrate natively with cloud platforms or offer tailored application support.

\subsection{Edge Gateway Design and Flow}
\begin{figure}
    \centering
    \includegraphics[width=0.7\linewidth]{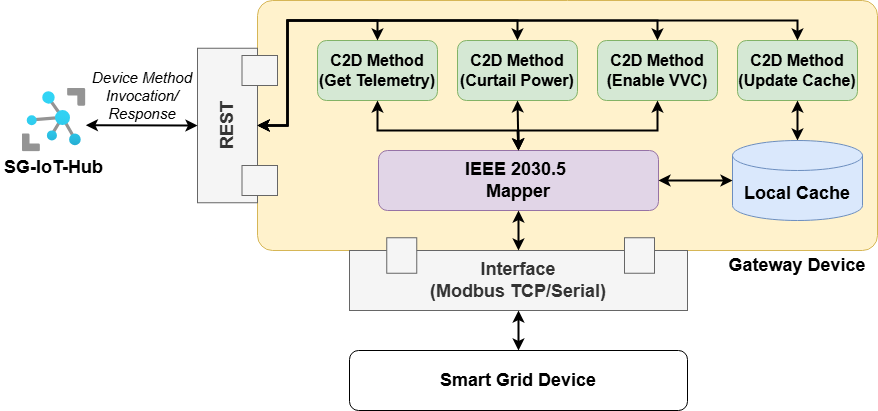}
    \caption{Device method invocation flow}
    \label{fig:device-method}
\end{figure}

At the core of the Smart Grid (SG) Gateway concept is an edge-level microservice architecture that enables flexible configuration and control delegation. Unlike monolithic systems, microservices are loosely coupled and independently deployable, and as such enhance robustness, scalability and maintainability. Each service encapsulates specific logic for telemetry, remote configuration and device control, supporting modular edge functionality. Gateway devices are orchestrated via a cloud-based SG management platform, which leverages cloud scalability to handle telemetry data, dispatch control signals, perform software updates and manage device mappings to IEEE 2030.5 and VVCs. This architecture offloads complex logic from the edge to the cloud, simplifying the management of numerous deployed gateways. Key cloud-enabled functions include SG device telemetry capture, DER curtailment, Local Volt-VAR Curve operation and updates to gateway cache for IEEE 2030.5 mappings and VVC bounds. Communication is facilitated using Azure Direct Methods, allowing device-specific actions to be triggered through an IoT Hub via REST compliant requests \cite{shi2019azure}. Figure \ref{fig:device-method} illustrates how device functions are invoked on a gateway device.

\noindent In order for a request to be valid, it must be authenticated through the cloud system before being invoked on a device or suite of devices. For example, a request for telemetry on a specific register through a gateway device would be made via POST request.

\begin{figure}
    \centering
    \includegraphics[width=1\linewidth]{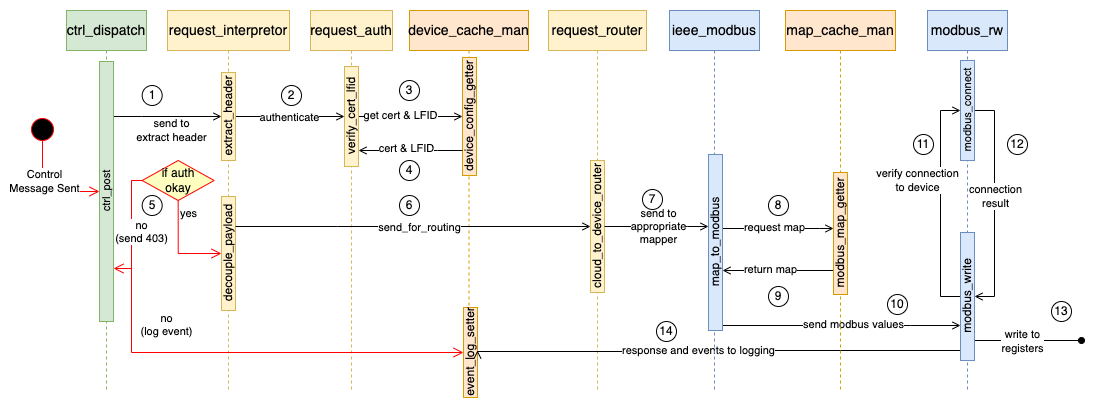}
    \caption{Cloud to device control signal data flow}
    \label{fig:receive-control}
\end{figure}

\noindent \textbf{Curtail Power}: The API endpoint defined for this action on the gateway is determined by specifying the device to be triggered (e.g., "gateway\_device\_XYZ") invoking the method "curtailPower" and passing a write command with a Modbus register and value, with a request made to this endpoint triggering the data flow present in \ref{fig:receive-control}. This command curtails active power on the connected SG device by a given percentage. These control signals are defined in the cloud system to be actuated on the connected SG device through the gateway.

\begin{figure}
    \centering
    \includegraphics[width=1\linewidth]{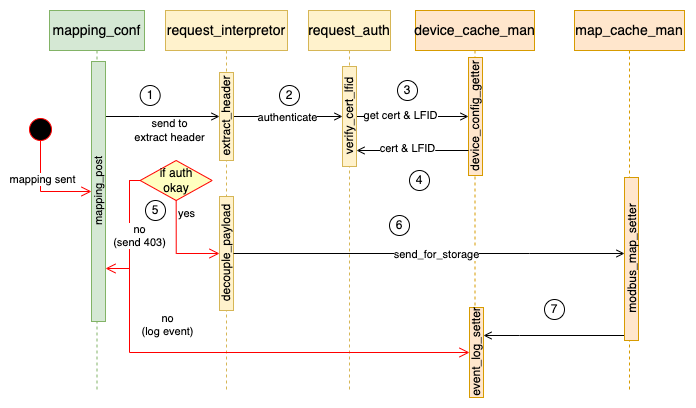}
    \caption{Cloud to device mapping data flow}
    \label{fig:receieve-mappings}
\end{figure}

\noindent\textbf{Update Gateway Cache}: Requests to this endpoint initiate the dataflow illustrated in Figure \ref{fig:receieve-mappings}. Mapping specifications transmitted from the cloud define how data from specific SG devices is aligned with standards such as IEEE 2030.5 for purposes including control decision-making, monitoring and historical data storage. To support dynamic updates, a local cache is maintained on each gateway device, offering high-performance, small-scale storage. Cache entries are structured as dictionaries and device methods enable management of this cache. For example, the "updateGatewayCache" function updates mappings by specifying a device type and associated mapping. Gateways are initially provisioned with a base mapping version, which can be seamlessly updated via the cloud to accommodate changes in IEEE 2030.5, thereby enhancing system adaptability and future-proofing. This mechanism also enables dynamic updates to the local Volt-VAR Curve, allowing refined control directly to the edge.

\begin{figure}
    \centering
    \includegraphics[width=1\linewidth]{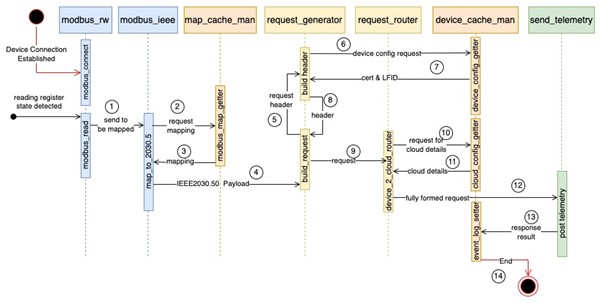}
    \caption{Device to cloud telemetry data flow}
    \label{fig:device-cloud-telemetry}
\end{figure}

\noindent \textbf{Device to Cloud Telemetry}: This dataflow governs device-to-cloud communication for capturing SG device telemetry and is initiated via the "getTelemetry" device method. Upon invocation, the gateway device retrieves specified values from the Modbus telemetry registers, maps them to corresponding IEEE 2030.5 entries, and transmits the data trhoguh a central message hub to the SG cloud platform for processing and storage. Figure  \ref{fig:device-cloud-telemetry} illustrates the sequence triggered by any valid telemetry value.

\subsection{Cloud Twin Implementation}
\begin{figure}
    \centering
    \includegraphics[width=0.8\linewidth]{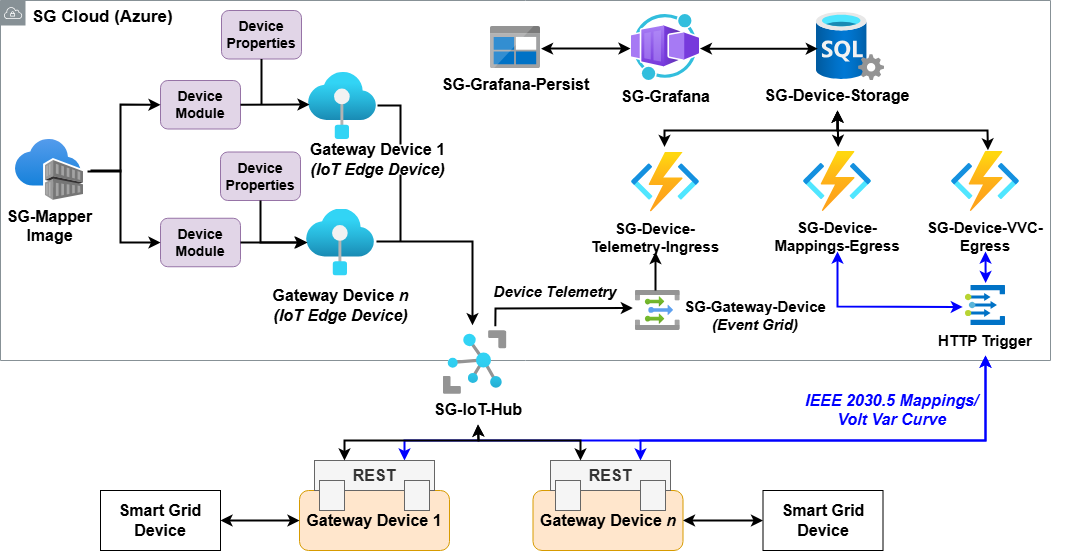}
    \caption{Cloud system architecture}
    \label{fig:cloud-arch}
\end{figure}

\noindent Within this section we outline the cloud architecture, built on Microsoft Azure's IaaS platform, to support gateway device management, data acquisition, control, processing, storage and visualisation. The system includes APIs for telemetry retrieval, control signal transmission and SG device mapping management, as illustrated in Figure \ref{fig:cloud-arch}.

\begin{figure}
    \centering
    \includegraphics[width=0.7\linewidth]{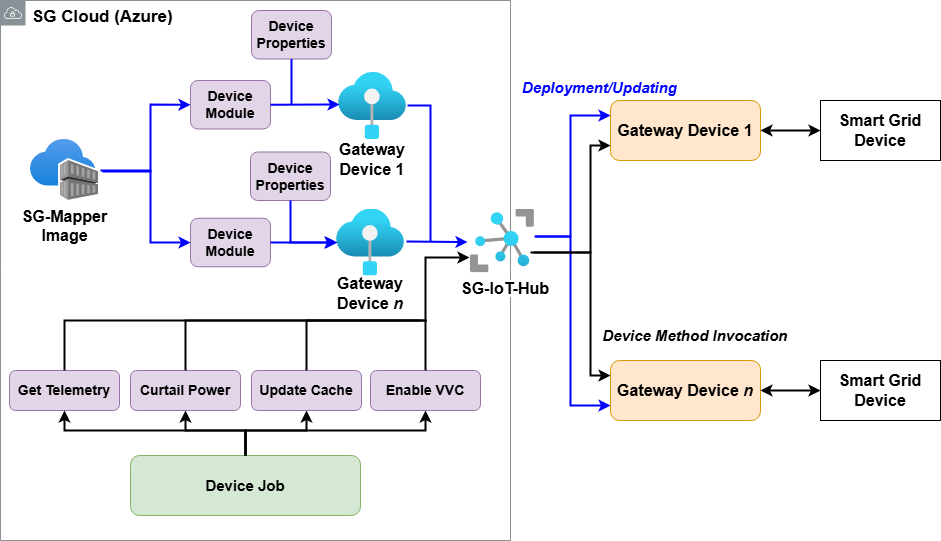}
    \caption{Gateway device deployment and scalability architecture}
    \label{fig:cloud-device-management}
\end{figure}

\noindent\textbf{IoT Hub}: Serves as the central access point, handling authentication, bidirectional data flow, device twin updates and IEEE 2030.5-compliant telemetry ingestion.
\noindent\textbf{Device-Telemetry-Ingress}: Is an Azure Function triggered by the Gateway-Device Event Grid to parse and store telemetry.
\noindent\textbf{Device-Mappings-Egress}: Responds to authenticated REST requests, delivering IEEE 2030.5-compliant mappings.
\noindent\textbf{Device-VVC-Egress}: Responds to authenticated REST requests, returning VVC bounds.
\noindent\textbf{Device-Storage}: Is a SQL-based repository for telemetry, gateway metadata, standard mappings and Volt-VAR Curve bounds.
\noindent\textbf{Dashboard-Service}: Leverages Grafana (deployed via Azure Container Apps) to visualise telemetry from Device-Storage.
\noindent\textbf{Dashboard-Service-Persist}: Retains Grafana settings such as dashboards and user accounts using an external SQLite database.
\noindent\textbf{Gateway-Registry}: Hosts container images for gateway device software across hardware variants via Docker.
\noindent\textbf{IoT Edge}: Provisions and monitors deployed edge devices, each with a Device Twin for configuration. The Device Module defines the deployed software, such as mapping containers, while Device Properties specific operational settings such as REST endpoint URLs, connected device types, IP addresses and communication modes. IoT Edge interaction with gateway devices in the field is described in Figure \ref{fig:cloud-device-management}.

\section{Results}

\noindent To support testing the SG gateway device and platform, a laboratory test environment was utilised consisting of a Fronius Primo single-phase inverter acting as the representative SG device. The smart grid gateway was implemented using a Raspberry Pi 4, configured to communicate with the inverter over SunSpec Modbus via wireless TCP/IP connection. This setup served as the basis for evaluating both mapping of telemetry and the dynamic control functionalities of the system. 

\subsection{Mapping Smart Inverter Telemetry to IEEE 2030.5}

In an effort to validate the SG platform's mapping layer, the laboratory setup was leveraged to gather telemetry data from the inverter (SG device) over TCP/IP through the gateway. This data included data points such as AC active power, AC voltage and reactive power, retrieved in real-time using SunSpec compliant Modbus registers. These data points were then mapped to IEEE 2030.5 compliant telemetry structures using the mapping service hosted on the gateway. A typical example is presented in Table \ref{tab:mapping}, where telemetry values from the Modbus interface are mapped to their respective IEEE 2030.5 fields. 

\begin{table}[h]
\centering
\caption{Example Mapping from SunSpec Modbus Registers to IEEE 2030.5 Telemetry Fields}
\label{tab:mapping}
\begin{tabular}{|c|c|c|c|}
\hline
\textbf{Parameter} & \textbf{SunSpec Modbus Register} & \textbf{IEEE 2030.5 Field Path} & \textbf{Units} \\
\hline
Active Power (W) & 40083 & \texttt{DERStatus\slash W} & Watts \\
\hline
AC Voltage & 40072 & \texttt{DERStatus\slash V} & Volts \\
\hline
AC Current & 40076 & \texttt{DERCapability\slash Amp} & Amperes \\
\hline
Frequency & 40070 & \texttt{DERStatus\slash Hz} & Hertz \\
\hline
Reactive Power & 40084 & \texttt{DERStatus\slash VAR} & VAR \\
\hline
\end{tabular}
\end{table}

\noindent A representative JSON payload produced by the gateway for cloud ingestion is presented in listing 1.1, which describes an IEEE 2030.5 mapping for an Active Power telemetry reading gathered from the inverter in real-time.

\begin{lstlisting}[language=json,caption={Example IEEE 2030.5 Reactive Power Reading}]
[
  {
    "40083": {
      "ReadingType": {
        "description": "W (Active power)",
        "accumulationBehaviour": 12,
        "commodity": 1,
        "dataQualifier": 0,
        "flowDirection": 1,
        "intervalLength": 3600,
        "kind": 0,
        "numberOfConsumptionBlocks": 0,
        "numberOfTouTiers": 0,
        "phase": 0,
        "powerOfTenMultiplier": 0,
        "tieredConsumptionBlocks": "false",
        "uom": 38
      },
      "Reading": {
        "consumptionBlock": "0 - N/A",
        "qualityFlags": "01",
        "timePeriod": {
          "duration": 60,
          "start": 1767148190
        },
        "touTier": "0 - N/A",
        "value": 1914
      }
    }
  }
]
\end{lstlisting}

\begin{figure}
    \centering
    \includegraphics[width=0.9\linewidth]{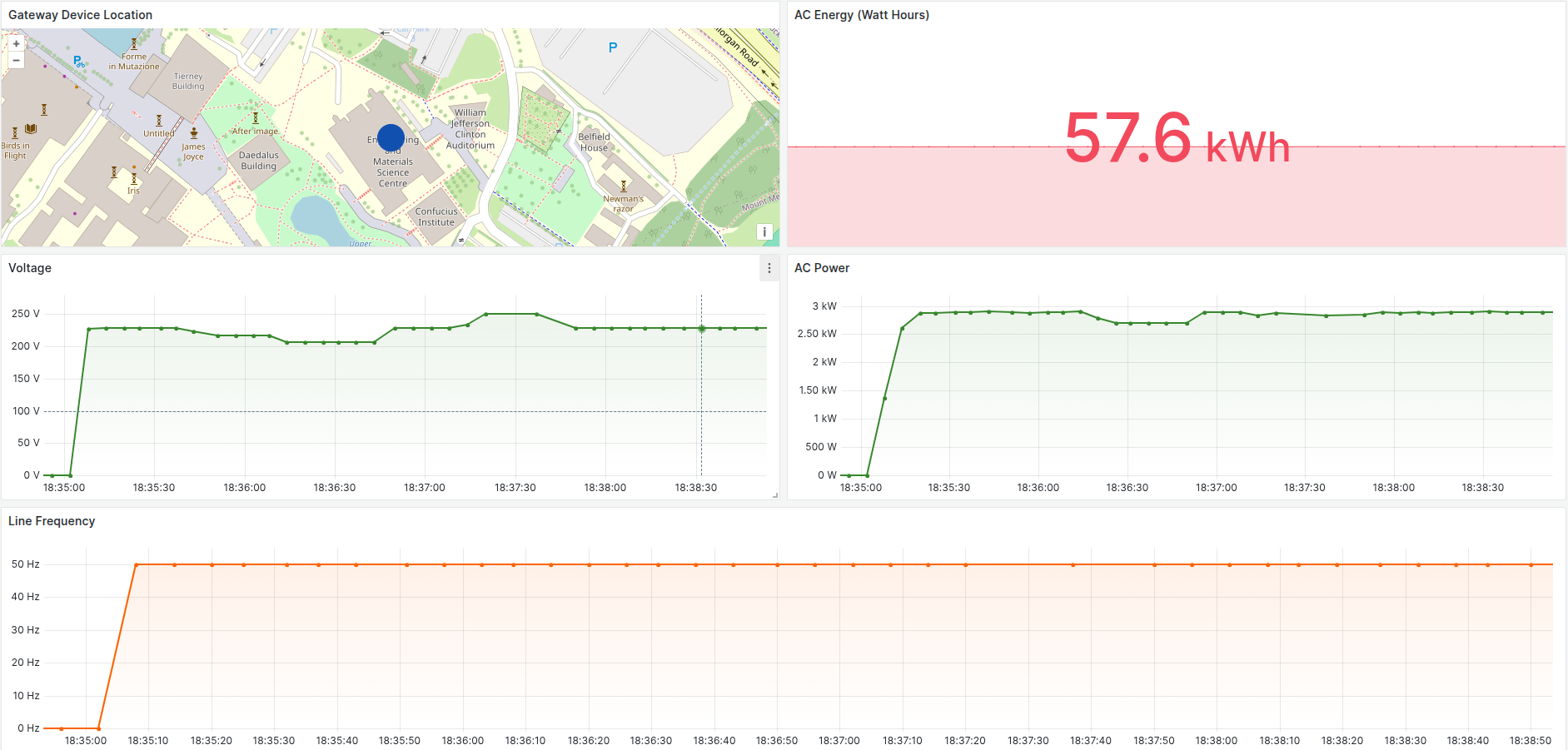}
    \caption{SG Platform visualisation dashboard}
    \label{fig:dashboard}
\end{figure}

\noindent Figure \ref{fig:dashboard} presents a sample of the SG Platform visualisation of gathered telemetry readings from the test system in the laboratory, including device location, AC Power, Voltage, AC Energy and Line Frequency viewable.

\subsection{Dynamic Volt-VAR Curve Update and Execution}

A key element of the work described in this paper was validating the ability to dynamically deploy and execute Volt-VAR Curves during real-time grid simulation. This test utilised the laboratory Hardware-in-the-Loop (HiL) setup to emulate changes in voltage conditions in a controlled manner, with real-time curve logic handled by the gateway device. Two separate Volt-VAR Curves were defined for the experiment to limit VAR injection, these curves were ingested by the gateway device in JSON format. Listing 1.2 describes the operational bounds of the first Volt-VAR Curve (VVC1) tested, while listing 1.3 describes the operational bounds of the second curve (VVC2) tested.

\begin{lstlisting}[language=json,caption={Volt-VAR Curve 1 Bounds}]
{"VOLTAGE_THRESHOLDS": {"V90": 207.0, "V95": 218.5, "V105": 241.5, "V110": 253.0}, "REACTIVE_POWER_LIMITS": {"Q_EXPORT": 45.4, "Q_ABSORB": -45.4, "UNITY": 0.0}, "ACTIVE_POWER_LIMIT": 75.0}
\end{lstlisting}

\begin{lstlisting}[language=json,caption={Volt-VAR Curve 2 Bounds}]
{"VOLTAGE_THRESHOLDS":{"V90":207.0,"V95":218.5,"V105":241.5,"V110":253.0},"REACTIVE_POWER_LIMITS":{"Q_EXPORT":40.5,"Q_ABSORB":-40.5,"UNITY":0.0},"ACTIVE_POWER_LIMIT":100.0}
\end{lstlisting}

\begin{figure}
    \centering
    \includegraphics[width=1\linewidth]{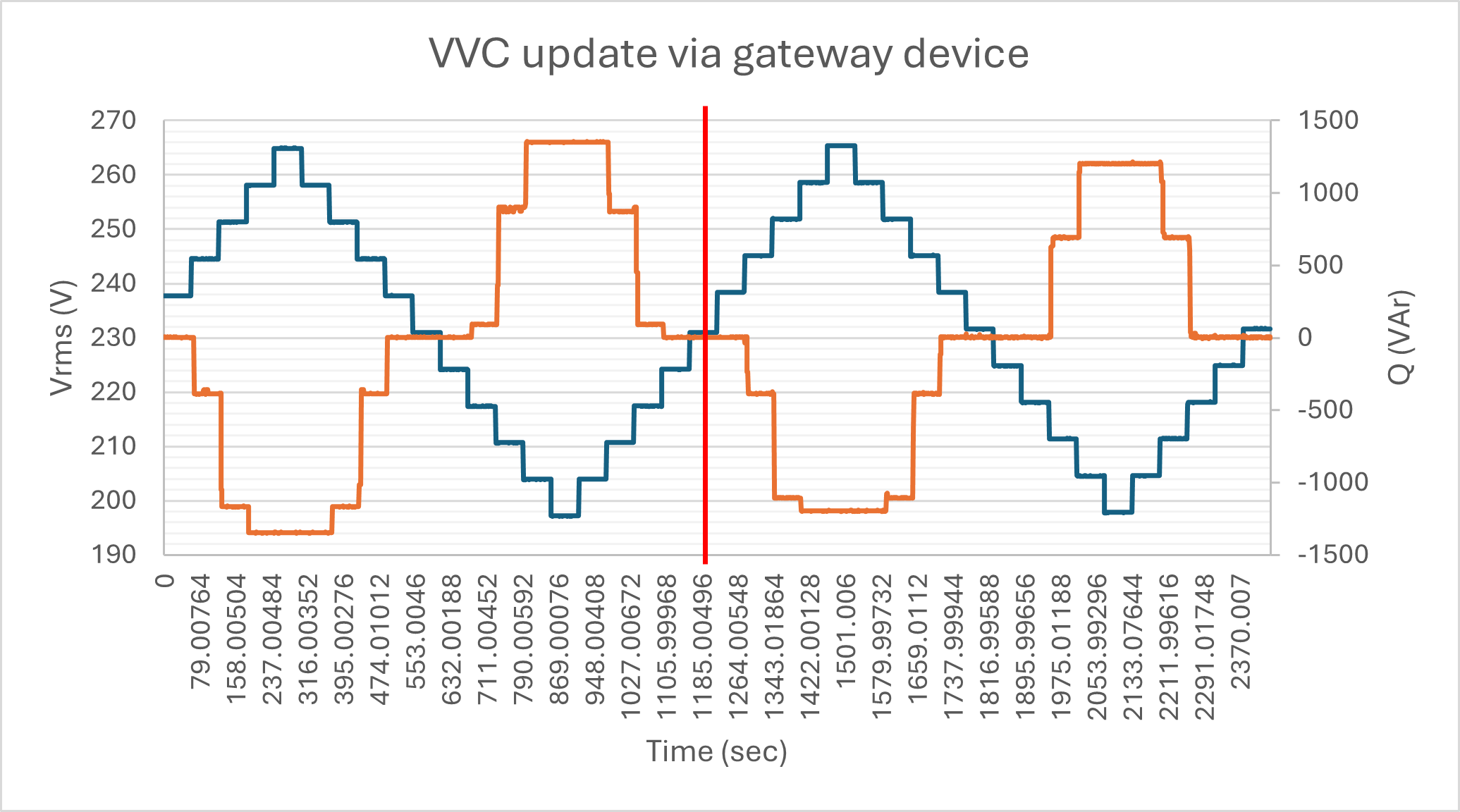}
    \caption{Reactive power response before and after VVC update}
    \label{fig:vvc_plot}
\end{figure}

\noindent Initially, VVC1 was active on the gateway, midway through the grid simulation, a new curve (VVC2) was dynamically pushed to the gateway device through using the "updateGatewayCache" method described in Section 4.2 (Edge Gateway Design and Flow). The curve data was passed from the cloud to the gateway, which translated used the JSON structure to generate reactive power setpoints which were acted upon the inverter through Modbus TCP/IP. Figure \ref{fig:vvc_plot} shows a time-series plot of voltage and reactive power across the testing period. The point of VVC updated is marked, and the transition in inverter behaviour is visible in the change in peaks and troughs between Q(Var) plots of both curves where reactive power output is reduced accordingly under VVC2 post-update, confirming the update had been successfully applied and executed. This test demonstrates the systems ability to dynamically modify grid-support behaviours in real-time and seamless communication from cloud to edge to device.

\newpage

\section{Future Work}
\begin{figure}[H]
    \centering
    \includegraphics[width=1\linewidth]{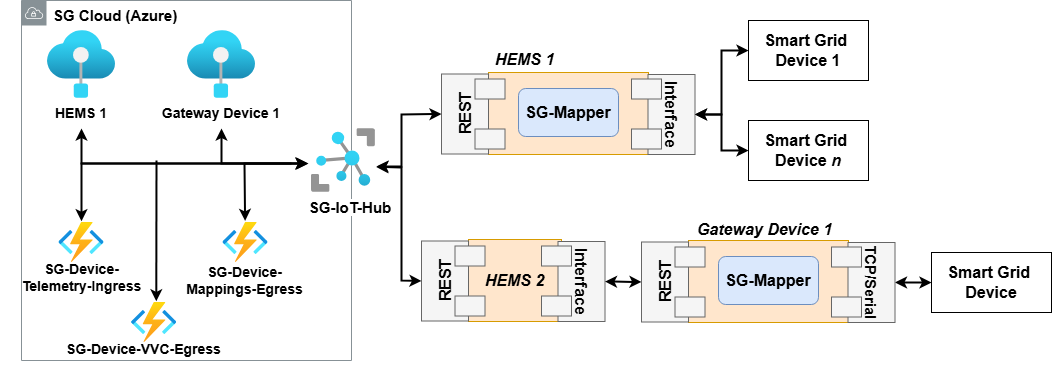}
    \caption{Cloud architecture to support HEMS/third-party cloud systems}
    \label{fig:cloud-arch-hems}
\end{figure}

\begin{figure}[H]
    \centering
    \includegraphics[width=0.7\linewidth]{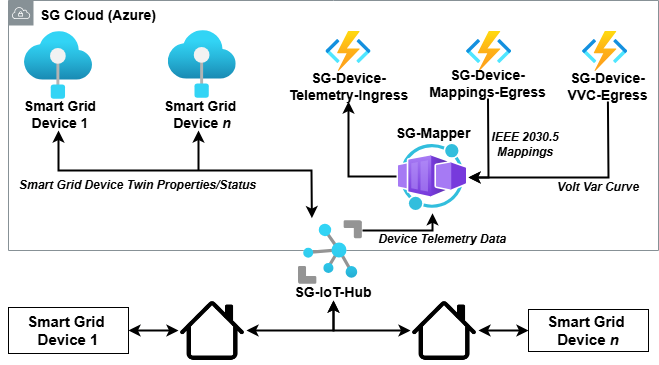}
    \caption{Cloud architecture to support native smart grid device control}
    \label{fig:cloud-arch-native}
\end{figure}

While the current cloud system assumes a gateway device sits between SG devices and the cloud, alternative architectures are emerging. Some buildings may use local energy management systems (HEMS) or third-party cloud platforms for SG device monitoring. In other cases, certain smart inverters support direct communication via interfaces like REST and such setups are currently being explored in regions like South Australia \cite{SA2023}. Given its modular, containerised design and loosely coupled services, the platform described in this paper can be adapted to support these alternative architectures with minimal refactoring. For instance, Figure \ref{fig:cloud-arch-hems} presents a scenario where the gateway mapper runs directly on a HEMS managed multiple SG or alternatively, a gateway may sit between. Figure \ref{fig:cloud-arch-native} outlines a cloud-centric model where native inverters connect directly to the network. In this case, no field gateway is required; the gateway mapper operates in the cloud, translating and processing IEEE 2030.5 mappings as needed per SG device. Supporting this paradigm would require some adjustments to the mapper software for multi-device support and scalability, but it demonstrates the system's architectural flexibility for future SG deployments.

\begin{credits}
\subsubsection{\ackname}  This study was funded by SEAI (Sustainable Energy Authority Of Ireland) in collaboration with ESB Networks, with support from SIRFN and OPENSVP.
\end{credits}
%
%
%
%

\end{document}